\documentclass[12pt,preprint]{aastex}

\shortauthors{Kafka et al.}

\begin{document}


\title{New Complexities in the Low-State line profiles of AM Herculis\footnote{Observations reported here were obtained at the MMT Observatory, a joint facility of the Smithsonian Institution and the University of Arizona.}}


\author{Kafka, S.\altaffilmark{1}}
\affil{Spitzer Science Center/Caltech, MC 220-6, 1200 E.California Blvd, Pasadena, CA 91125, USA}

\author{Ribeiro, T., Baptista, R}
\affil{Departamento de Fisica, Universidade Federal de Santa Catarina, Campus Trindade, 88040-900 FlorianÃ³polis, SC, Brazil}

\author{Honeycutt, R.K.}
\affil{Indiana University, Astronomy Department, Swain Hall West, Bloomington, IN 47405, USA}

\and

\author{Robertson, J.W.}
\affil{Department of Physical Sciences, Arkansas Tech University, Russellville, AR 72801-2222, USA}


\altaffiltext{1}{email: stella@ipac.caltech.edu}


\begin{abstract}

When accretion temporarily ceases in the polar AM Her, the emission line profiles are known to develop several distinct components, whose origin remains poorly understood. The new low-state spectra reported here have a more favorable combination of spectral resolution (R$\sim$4500), time resolution ($\sim$3-min exposures), and S/N than earlier work, revealing additional details of the orbital dependence of the line profiles. The central strong feature of H$\alpha$ is found to be composed of two components of similar strength, one having K$\sim$100 km sec$^{-1}$ and phased with the motion of the secondary star, the other having little or no detectable radial velocity variations. We attribute the central line component to gas near the coupling region, perhaps with a contribution from irradiation of the secondary star. The two satellite components have RV offsets of $\sim\pm$250 km sec$^{-1}$ on either side of the central strong H$\alpha$ peak.  These satellites most likely arise in large loops of magnetically confined gas near the secondary star due to magnetic activity on the donor star and/or interactions of the magnetic fields of the two stars. Doppler maps show that these two satellite features have concentrations at velocities that match the velocity locations of L4 and L5 in the system.

\end{abstract}


\section{Introduction}

In Cataclysmic Variables (CVs)\footnote{CVs are semi-detached binaries in which a white dwarf is accreting material from its Roche Lobe filling low mass companion.} with orbital periods longer than 3 hours, chromospheric activity on the mass donor, low MS star is considered to be the principal mechanism extracting angular momentum from the system, thus driving its evolution. Conclusive observations on the presence and characteristics of activity on this star are difficult to extract, due to the presence of accretion that hides or mimics common activity indicators (such as x-ray and H$\alpha$ emission). Recent observations of the magnetic CV prototype AM Her in its extended 2003-2005 low state revealed multi-component H$\alpha$ emission lines whose properties strongly suggest that they arise from large magnetic loops surrounding the secondary star (Kafka et al. 2005a; 2006; paper I and II respectively). The longevity of those structures suggested magnetic interaction between the magnetic WD and its companion for their origin. We have interpreted these features as large semi-permanent loop prominences having magnetically-confined gas motions. Regardless of their origin, those structures provided a well-needed spectroscopic evidence for magnetic activity on the donor star in such systems, potentially providing an important new tool for studying braking on a magnetically-coupled stellar wind on the donor star at times of negligible accretion.

The spectra reported in Papers I and II suggested that higher spectral and time resolution data, could reveal hidden information on the structure and kinematics of the low-state line components in AM Her providing further clues to the nature of polars in the low-state. To that end we obtained $\sim$2 orbits of low-state AM Her\footnote{A review of the history of low-state spectra of AM Her are presented in papers I and II and will not be repeated here} spectra using the MMT on 2006-Sep-16, whose analysis is reported here. Section 2 provides details regarding the observations and reductions, Section 3 discusses the results, and our conclusions are summarized in Section 4.

\section{Spectroscopic Observations}

Our new spectroscopic monitoring data were obtained during one night in 2006-Sept-16 UT, when AM Her was in a low optical state (V$\sim$15.5), using the 1200 line grating of the Blue Channel MMT Spectrograph. This yielded resolution of 1.4$\AA$ (full width at half maximum) effectively covering from $\sim$6000-7200$\AA$. Exposures times were mostly 200 sec and a  wavelength calibration spectrum of a  HeNeAr lamp was obtained every 10 exposures. Taking into account comparison spectra and readout time, a total of $\sim$70 exposures were acquired over $\sim$4.3 hr, under clear skies. For data analysis we used the twodspec/onedspec packages of IRAF\footnote{IRAF is distributed by the National Optical Astronomy Observatories, which are operated by the Association of Universities for Research in Astronomy, Inc., under cooperative agreement with the National Science Foundation} and custom programs (which will be discussed later).

\section{Discussion}

\begin{figure}
\epsscale{1.0}
\plotone{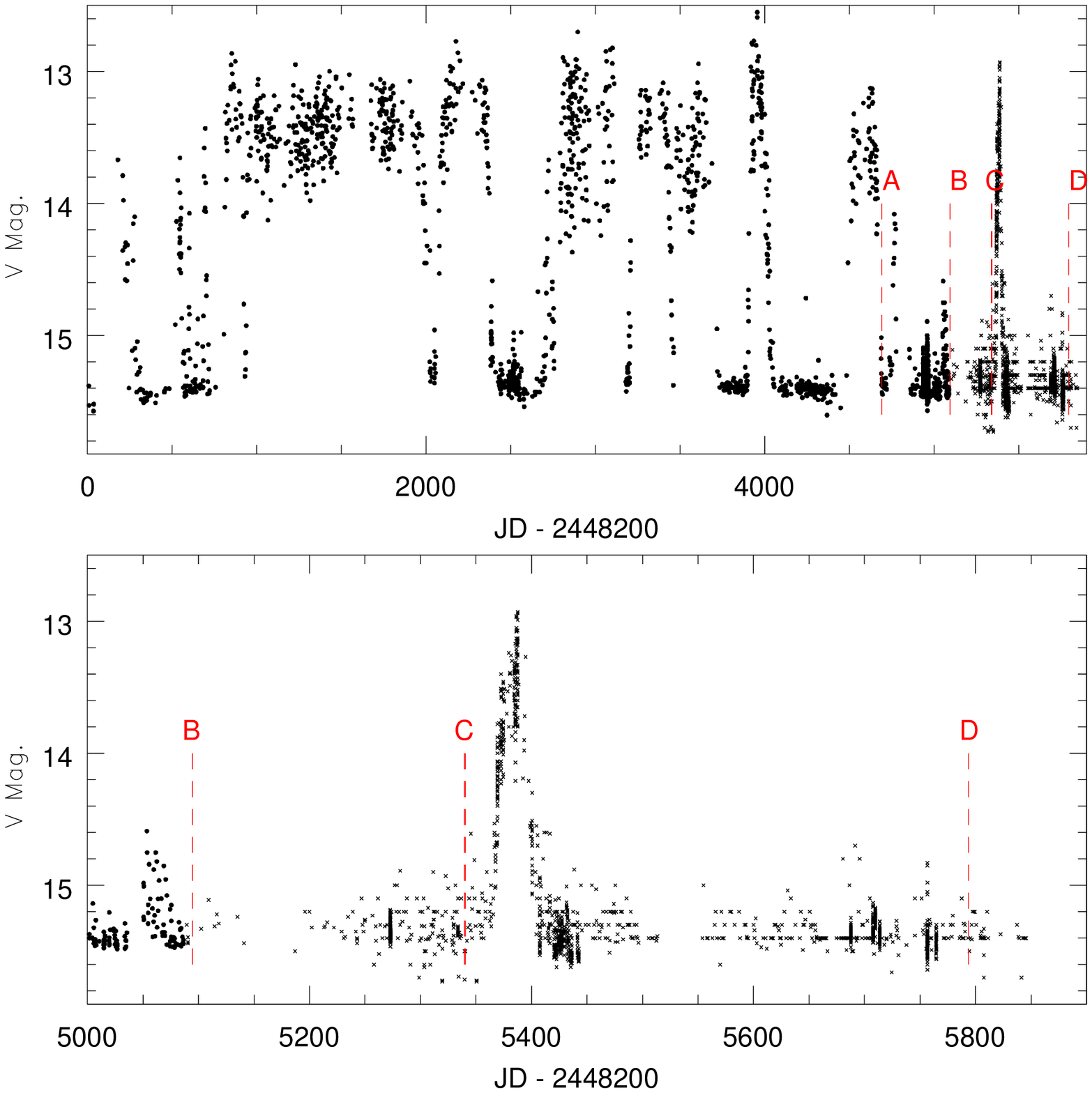}
\caption{RoboScope (closed circles) and AAVSO (crosses) long term light curve of AM Her with all the dates of our spectroscopic observations marked, including those from Papers I and II. 1: KPNO 2.1m (Sept 2003); 2,3: WIYN 3.5-m ( Feb 2005; June 2005); 4: MMT 6.5-m (Sept 2006). A short high state preceded our MMT spectroscopic observations, after which the system dropped again to the same low-state brightness.\label{lc}}
\end{figure}

\begin{figure}
\epsscale{1.0}
\plotone{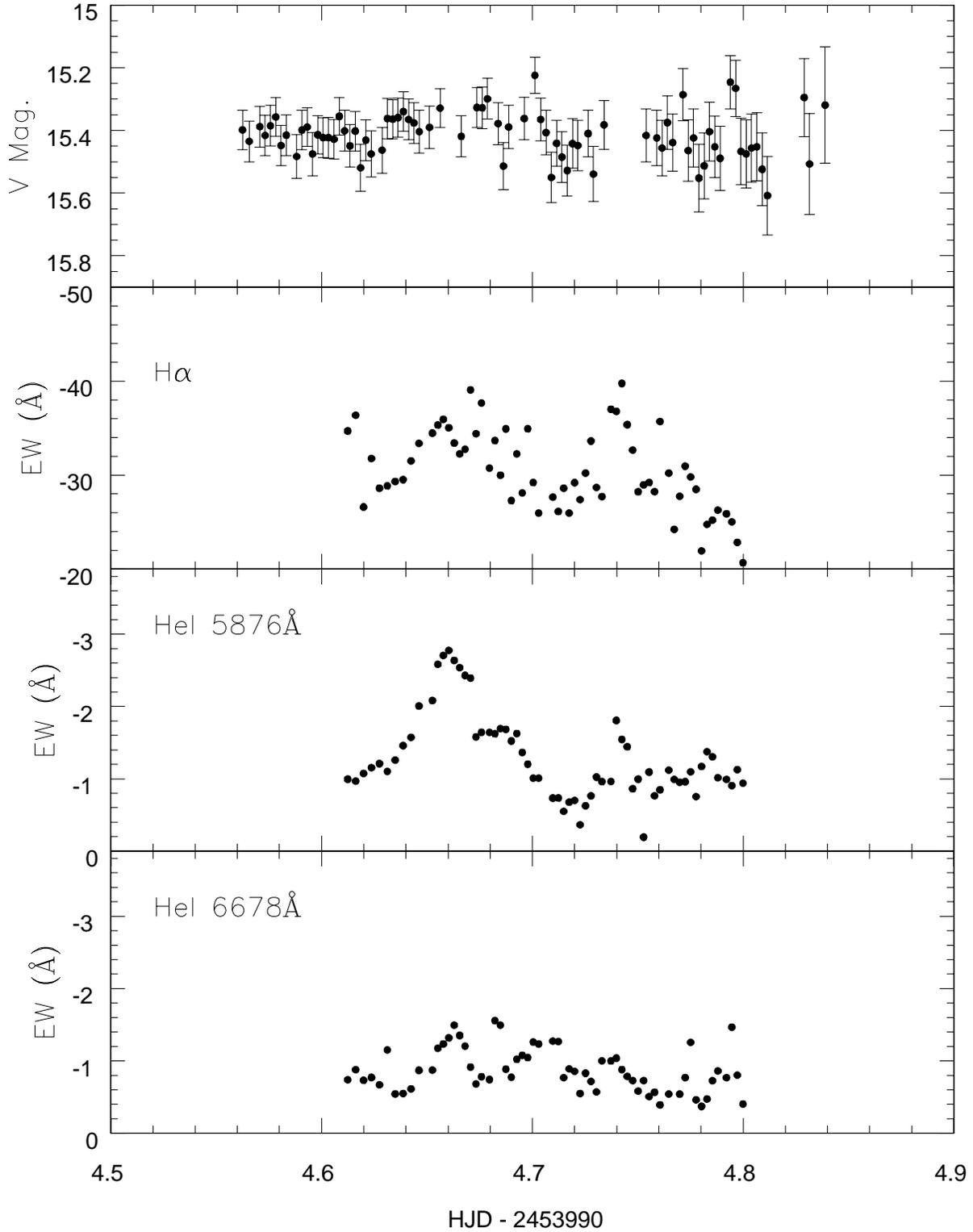}
\caption{AM Her light curve at the time of our spectroscopic observing runs and EWs of the main emission lines present. See text for details.\label{lcEWs}}
\end{figure}

\begin{figure}
\epsscale{1.0}
\plotone{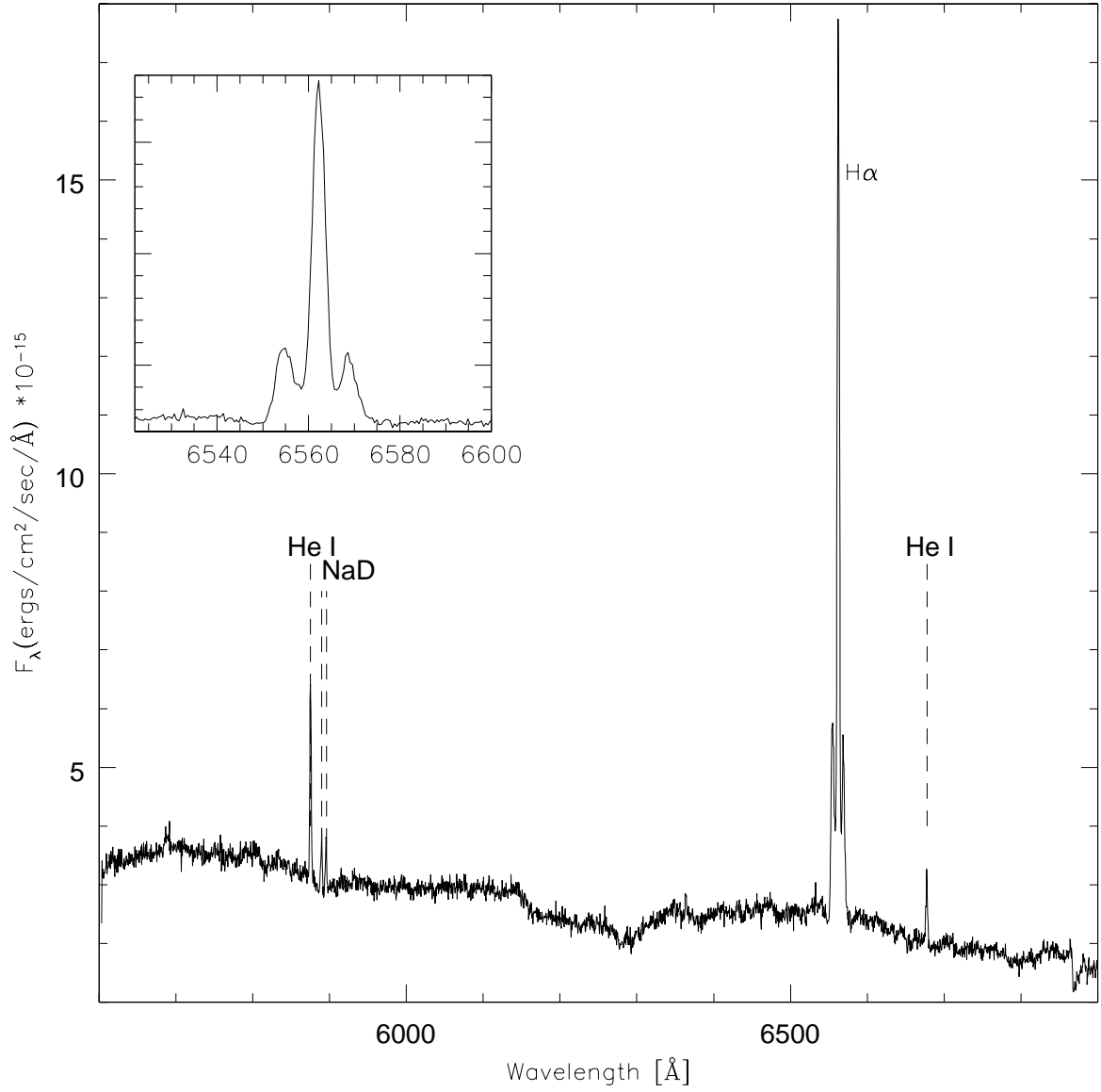}
\caption{Example spectrum of AM Her; in the inset we present a zoom-in of the H$\alpha$ line.\label{example}}
\end{figure}

Figure~\ref{lc} shows the 1990-2005 RoboScope\footnote{RoboScope, a 0.41-m telescope in Indiana, equipped for automated, unattended differential CCD stellar photometry (Honeycutt \& Turner 1992). AM Her has been in the RoboScope database since 1990; the resultant long-term light curve of the system is presented in Kafka et al. (2005). RoboScope was in a maintenance mode during 2006, therefore we complement its long-term light curve with a light curve from the AAVSO database.} light curve of AM Her complemented by the AAVSO light curve of the system. The times of the 2003-2005 observations from papers I and II are included in the figure (A-C), along with the time of the new MMT observations (D). As described in Kafka et al. (2005), Paper I and Paper II, AM Her in the
   low state has several kinds of photometric behavior.  The quiescence
   magnitude seems to always be V$\sim$15.5; however, sometimes 0.5 mag
   events are superimposed, while at other times a weak sinusoidal modulation
   is present.  On some occasions the system is particularly quiet with 
   no photometric modulation at all being apparent. The latter behavior best
   describes AM Her during our MMT spectroscopy, as seen from the simultaneous
   photometry in the top panel of Figure~\ref{lcEWs}. These data were obtained using
   the 0.3-m telescope at the ATU observatory, and were reduced in the
   manner described for the ATU data in Kafka et al. (2005).  Although
   the accuracy deteriorates near the end of the sequence, it appears that
   accretion in AM Her was very low or absent during the times of our
   spectroscopy.  The bottom panels of Figure 2 show that the strengths 
   of the emission lines varied considerable during this time, particularly
   He I 5876$\AA$. Figure~\ref{example} presents a typical MMT spectrum of AM Her with the main features labeled. H$\alpha$ is in emission as are the HeI 5876$\AA$ and 6678$\AA$ lines. 

\begin{figure}
\epsscale{1.0}
\plotone{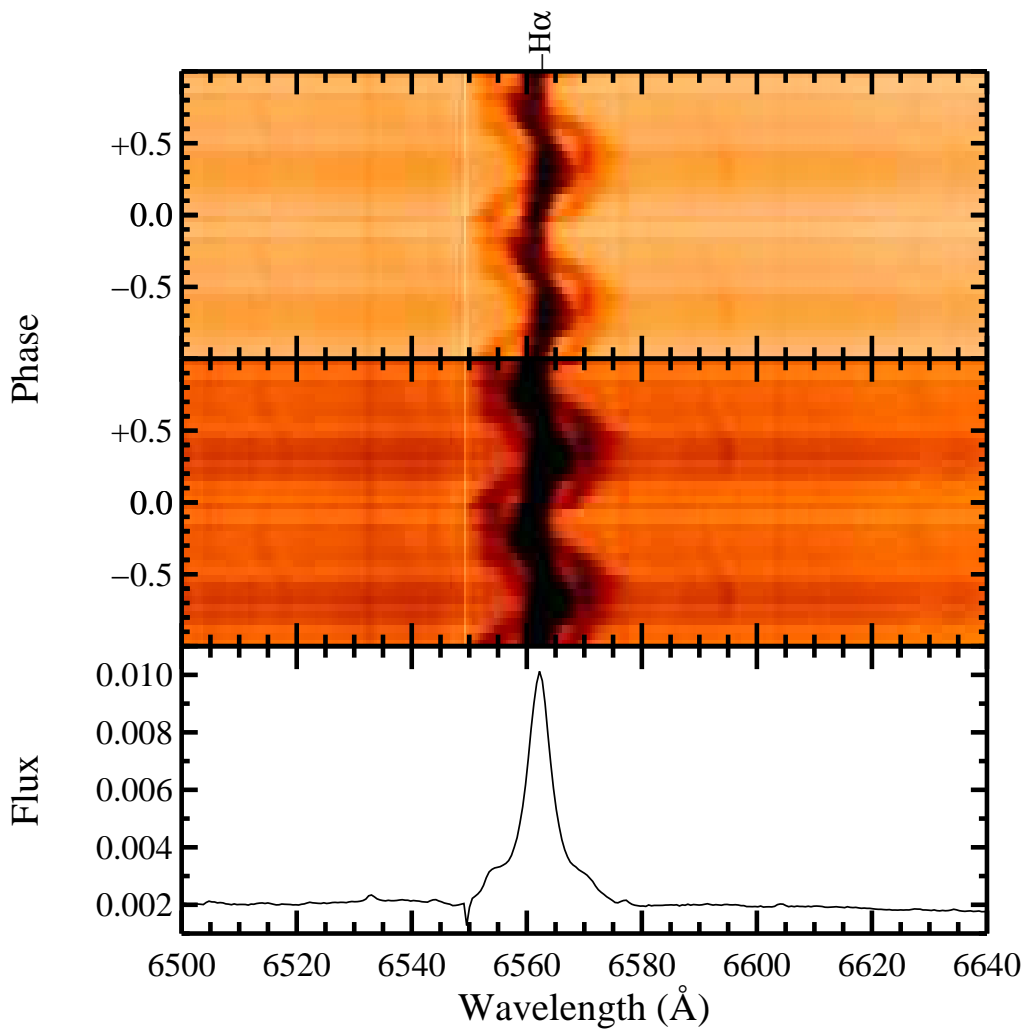}
\caption{Trailed spectrum of H$\alpha$; the line profile in the bottom
   panel is averaged over phases. The phases in the top two panels are
   repeated to provide two complete cycles. The central peak has two 
   components, a nearly stationary feature plus a low amplitude sine-like 
   curve. The latter has phasing similar to that of the secondary star,
   whose spectral features are visible in the middle panel.  The high velocity
   satellites appear as weaker sine-like curves which are difficult to
   trace at some phases where they merge with the central component and
   (perhaps) with each other; these satellites also have phasing similar
   to the donor star. Notice that the blocked column on the blue side of the H$\alpha$ line does not interfere with its blue emission wing.
\label{trailedHa}}
\end{figure}

The NaD lines are also in emission; extra care is taken during the analysis of these lines because of contributions from Tucson city lights. The inset in figure~\ref{example} presents an expanded view of the H$\alpha$ line. The central peak is accompanied by two satellites similar to those discussed in Paper I.  These satellites have been present in all our low-state spectroscopic runs, and have had similar strengths, 
   radial velocity amplitudes, and phasings for each run.  However, the
   satellites are much better defined in this current data because of the
   higher spectral resolution, time resolution, and S/N. For the analysis of the data in this work, we used the spectroscopic ephemeris of paper I\footnote{T$_{0}$=HJD 2,446,603.403(5) + 0.12892704(1)E}, defining orbital phase zero at inferior conjunction of the secondary star. Using this ephemeris we constructed the trailed spectrogram of figure~\ref{trailedHa}; the strong features are emphasized in the top panel and the weaker features in the middle panel.

The top panel clearly indicates substructure in the central peak. (This central peak appears single in Papers I and II, due to lower spectral resolution and S/N.)  There is a nearly stationary component, and another which appears sinusoidal with K$\sim$100 km sec$^{-1}$. This sinusoid is phased with the radial velocity of the absorption lines from the secondary star. and may arise on the irradiated portion of the donor star near L1.  The K amplitude of the tip of the Roche lobe of the secondary star\footnote{see the Doppler maps of Figs 9 and 10 which display the orbital velocities on the surface of the secondary star. A relevant discussion of the system parameters used is presented in section 3.1 and table 2} is $\sim$150 km sec$^{-1}$. Other portions of the "neck" of the Roche lobe have even larger orbital velocity, encouraging consideration of other origins for the sinusoidal component of the central peak. We will return to discussion of the two components of the central peak later in the paper.

A different view of the satellites appears in the middle panel of figure~\ref{trailedHa}. There are two main satellites visible at most phases. The satellites
    become very complex between phases 0.2 and 0.45, perhaps becoming three
    features.  In fact, the strong central component of H$\alpha$ also becomes
    double in this phase interval, helping provide the impression (discussed
    earlier) that this central component is composed of a stationary component
    plus a sinusoid.  This complex behavior makes it very difficult to tell
    if the overall phase dependence of the satellites is best described by
    two crossing sinusoids, or by portions of two non-crossing sinusoids.

 To better study the kinematic behavior of the H$\alpha$ line, we used a three Gaussian fit decomposition. (Note that the central H$\alpha$ component has not been deblended because its components (shown in the trailed spectrogram) are poorly resolved; therefore the RV measured is determined from the fit of a single Gaussian fit.) The line profiles are clearly more complex than 3 Gaussians (particularly in the phase interval 0.2-0.45); nevertheless this
     initial approach is an appropriate attempt to characterize the
    overall phase behavior of the satellites. Our denser time coverage and higher spectral resolution
    compared to Papers I and II allows tracing the RV curves of the satellites
    nearer to the central component than before. 

 Using sinusoids of the form {\it v(t) = $\gamma$ - Ksin[2$\pi$($\phi$ - $\phi_{0}$)]} we derived the systemic velocity ($\gamma$), the semi-amplitude velocity of the fit (K) and the zero phase ($\phi_{0}$) values, which are presented in table~1. Figure~\ref{rvfits} (top) presents the RV of the three main components of the H$\alpha$ line, with a sinusoidal fit superimposed on the RV of the central component. (The small phase offset of the fits in table~1 is due to the multiple components of the H$\alpha$ line which contribute to the line differently at each phase).  To first order, the satellites have similar behavior to the ones presented in figure~9 of paper I and in figure~3 of paper II. Characterizing the RV curves of the two main components of the satellites is challenging: one possible interpretation is that they represent two partial loops of streaming gas corotating with the donor star, in agreement with the conclusions of papers I and II. In the middle panel of figure~\ref{rvfits} we add the fit to the satellite radial velocity using the parameters from paper I. The bottom panel of Figure~\ref{rvfits} shows an alternative characterization of the RV curve of the satellites (green and blue fits), consisting of two crossing sinusoids\footnote{The obvious deviations of the data from the sinusoidal fits in the satellite RVs indicate that the satellites are not perfect sinusoids. However those fits are the best first approximation to the observed RVs.}; the parameters of the fits are included in table~1. 

 Note that the red satellite RVs do not match ${\it either}$ characterization over the phases 0.65-0.9.  Clearly our fits are a simplified representation and the satellites likely have contributions of two or more emission regions, each of whose visibility varies with phase due to some combination of obscuration and optical depth effects. A possible explanation could be that the satellites represent a collimated flow (eg. jet) similar to the jets of symbiotic stars (e.g. SS433; Gies et al. 2002). However, in AM Her, the spatial and secular stability of the satellites make this explanation highly improbable, since collimated outflows tend to evolve with time. 

 Trailed spectra of the underlying TiO bands are present in the continuum of the middle panel of figure~\ref{trailedHa} demonstrating that all features are following the motion of the donor star. The NaD doublet has been traditionally used to trace the motion of the donor star and map its irradiated inner hemisphere, therefore the K and $\gamma$ velocity of these two lines should be representative of the motion of the donor star. A trailed spectrum of NaD lines are presented in figure~\ref{trailedHeNaDHa}. There are
two main features clearly visible on the trailed spectrum, a component
that follows the motion of the secondary star and a puzzling
stationary component which remains after application of standard IRAF sky subtraction routines. Considering the 
position of AM Her (and the telescope) at the night of our observations, this component is expected to be due to strong contribution from the Tucson city lights. This stationary component is better seen at quadratures and is not taken into account in
the RV measurement of the line. We attempted to subtract the stationary component assuming
that it is represented by a Gaussian (which is
defined by the line prole around phases 0.25 and 0.75 where one can
see two peaks in the line). Unfortunately, the two components of each NaD line (aka, star+Tucson lights) are badly blended, resulting in
residuals even after subtraction at all phases. 
Such residuals appear in
our doppler maps (presented later in the paper) but do not afect the
fits to the radial velocities for the NaD lines. Therefore we investigated only the 
sinusoidal component. The possibility that the 
stationary component is intrinsic to the stellar spectrum is 
unlikely since such emission is not present in the NaD line of our AM Her low-state spectra taken at other epochs.

The K and $\gamma$ velocities for the HeI and NaD lines are included in table~1. Almost all lines have the same systemic velocity, although there is a notable 0.05 phase shift between the central H$\alpha$ line and its satellites (which is also clearly seen in the trailed spectrogram of figure\ref{trailedHa}). Comparing the K velocity of NaD with that of the HeI lines and the central component of H$\alpha$  we deduce that, to first order, neither lines originate from L1. In particular, both HeI emission lines appear to originate from a site between the L1 point and the center of mass of the system. We will discuss their origin in an upcoming section.

In figure~\ref{lcEWs} we present the equivalent widths of the HeI and H$\alpha$ emission lines with respect to time. Variations in the line strengths are present in all emission components, however they do not seem to be correlated with the brightness of the system (which is almost constant - within error). Also we do not detect any coherent orbital variations in the EWs of any line, nor in the light curve of AM Her at the times of the relevant spectroscopic observations.

\subsection{Doppler maps}

 Doppler tomography is a technique that was initially applied astronomically to create velocity maps for disk CVs (Marsh $\&$ Horne 1988; Kaitchuck et al. 1994) and it is slowly becoming popular for the study of other accreting sources such as Algols (eg. Richards 2004) and magnetic CVs (eg. Schwope et al. 1997). For the latter systems, the presence of the strongly irradiated inner hemisphere of the donor star, the ballistic/blobby accretion stream, and accretion toward one or two poles out of the orbital plane introduce additional parameters that are in need for careful modeling and understanding before they are securely identified in the (Vx,Vy) plane. This is still a subject of experimentation; therefore interpretation of Doppler maps in polars need to be made with additional caution. High resolution spectra of ionized HeII lines (such as HeII4686$\AA$ and 8236$\AA$) are used to trace parts of the system affected by the irradiating field of the heated magnetic pole(s), whereas bright Balmer (H$\beta$ and H$\gamma$) and HeI lines are used to primarily locate parts of the system associated with the accretion stream. A review of the magnetic CV doppler map characteristics for one and two pole accreting magnetic CVs can be found in Schwope et al. (1999); collections of tomograms are abundant in the literature (eg Hoard 1999). The main emission regions for the observed structures consist of the irradiated inner hemisphere of the donor star, the ballistic accretion stream (from L1) and the magnetically controlled part of the accretion. With the center of mass of the CV at the map center (Vx,Vy=0,0) the three components are concentrated in the left quadrants of the maps, with the majority of the ballistic stream on the -Vx,+Vy quadrant and the magnetically controlled trajectory toward the low-left (-Vx,-Vy) quadrant. Differences between observations and derived models (eg. Heerlein et al. 1999) are attributed to the complex accretion pattern and topology of the magnetosphere of the white dwarf. 

For AM Her only a few attempts have been made to construct doppler maps. Staude et al. (2004) present high state spectra of AM Her indicating that the observed accretion patterns may violate the canonical model of Roche-lobe overflow accretion from magnetic CVs. The observed H$\beta$ line consist of a bright narrow component due to irradiation of the inner hemisphere of the donor star, accompanied with a broad, diffuse component due to the accretion stream, primarily concentrated in the low-left quadrant of the tomogram. The structure of the trajectory in the HeII 4686$\AA$ line deviates from what is expected for the motion of the accretion stream, indicating that there is significant contribution in the z-direction (vertical to the orbital plane). Staude et al. (2004) propose an alternative scenario for the origin of this vertical component, which involves material ejected from the donor star via a prominence (as opposed to Roche lobe overflow) immediately coupling onto the magnetic field lines of the white dwarf. This unconventional accretion scheme reproduces the observed tomograms, confirming that accretion onto a magnetic WD is more complex.
Papadimitriou \& Harlaftis (2004) present H$\alpha$ trailed spectra of AM Her during four epochs of observations, in which a high velocity component is accompanying the central broad low velocity S-wave. The relevant doppler maps indicate that the majority of the emission is not following the trajectory expected from the ballistic stream; it rather appears as a ``bright blob'' starting on the L1 point and spreading toward the -Vy axis. In their high-state data an additional weaker ``spot'' is  accompanying the principal emission; this was attributed to accretion onto a second magnetic pole.

For our doppler maps we used the Doppler Tomography code designed by Spruit (1998). The assumptions of the code are standard for Doppler tomography and use the Maximum-Entropy (ME) reconstruction method of Lucy (1994). In short, the method is finding the minimum of the equation $Q = H + \alpha S$, which relates the likelihood of the fit ($H$) to the entropy of the map ($S$) for a given convergence factor ($\alpha$). The gas is assumed to be optically thin, and moving in the orbital plane. The entropy is calculated in relation to a gaussian hump floating
 default map. The resulting maps for the H$\alpha$ line are presented in figure~\ref{dopmapHa}. The center of mass of the system is at Vx,Vy=0,0 and the positions of the donor star and the white dwarf are marked. A solid line indicates the orbital velocity of gas falling along the ballistic stream trajectory, whereas the dashed line shows the corresponding Keplerian velocity at various points along the accretion stream. The contour lines delineate equal intensity surfaces of the grayscale of the map are marked at the top of each plot. The model parameters used are presented in table~2. Harrison et al. (2005) concluded that the donor stars in magnetic CVs are similar to single lower MS stars. The secondary star in AM Her is an M5$\pm$0.5 dwarf (Kafka et al. 2005a), which corresponds to a main sequence mass of 0.2M$_{\sun}$ (Allen, 2000). This is the value used for the mass of the donor star in our model.

 The Doppler maps of AM Her introduce puzzling new elements of the system. There are three major emission regions in the map corresponding to the central line component and the two satellites, and they are discussed separately in the following sections.

\subsubsection{Central (low velocity) Components} 

The right panel of figure~\ref{dopmapHa} demonstrates that the central emission in H$\alpha$ appears as an elongated structure at a range of velocities  spreading from the center of mass of AM Her (responsible for the stationary part of the central component) up to velocities close to that of the L1 point (responsible for the sinusoidal part of the central component). A similar feature is often seen in polars and often attributed to the irradiated inner hemisphere of the secondary star, but that does not appear to be the case here. The high-state
   velocity maps of AM Her of Staude et al. (2004) show a concentration very near L1, for both HeII and H$\beta$.  The velocity maps of
   AM Her by Papadimitriou $\&$ Harlaftis (2004) show a similar concentration but situated nearer to the center of mass than to L1. 

Since AM Her is not an eclipsing CV, we can not assess the vertical extent of the emitting source of the low velocity component (the component that lies perpendicular to the orbital plane). In general, though, there are two possible origins for its presence in the system. A stationary component with respect to the center of mass of the CV could result from a circumbinary disk. In a recent study of a number of magnetic CVs in the near-IR  five out of the six systems in the survey were found to have near-IR (up to 8$\mu$m) excess likely due to the presence of a circumbinary disk (Brinkworth et al. 2007). AM Her was not one of the systems discussed; however, existing Spitzer data suggest that there is no large near-IR excess in this system down to 8$\mu$m in the low state (Hoard et al. in preparation). 

Therefore, circumbinary gas is not responsible for the H$\alpha$ emission.  An indication of the origin of the stationary component can be found in the trailed spectra of the HeI lines (figures~\ref{dopmapHeI5876} and \ref{dopmapHeI6678}). We expected the HeI lines to trace parts of the donor star of the system (irradiated inner hemisphere). Instead, the majority of the emission seems to originate in a source near the center of mass, with the stationary component mapping the center of mass and a weak sinusoidal component leading to an elongation towards the velocity of L1 in the Doppler map. Visual inspection of the raw spectra eliminated the possibility that the stationary emission (in all HeI and H$\alpha$ lines) is due to an extended source in the foreground/background of the object. Furthermore, night sky and airglow emission does not appear in the HeI nor H$\alpha$ lines, ruling contamination from artificial sources as the source of the low velocity and stationary components. Therefore, it is logical to explore the second option for their origin, to assess the part of the system giving rise to the central components.

Gaensicke et al. (2006) calculate a low-state mass accretion rate of $\dot{M}$$\sim$6$\times$10$^{-12}$M$_{\sun}$yr$^{-1}$ which can be only due to wind accretion form the donor (as opposed to Roche-lobe overflow accretion). With such low $\dot{M}$, the accreted material should couple onto the white dwarf magnetosphere at a point closer to the donor star, likely resulting on a different (than in the high state) impact spot on the surface of the white dwarf.  Schwope et al. (2001) in a multi-epoch study of the eclipsing magnetic CV HU Aquarii noticed changes in the longitude of the accretion spot in the low state of the system with respect to the position of the spot in the high state. Those changes were a consequence of lower mass accretion rate in the low state leading to a move of the threading region closer to the donor star (and L1) than in the high state. This is a natural consequence of the balance of the ram pressure with the magnetic pressure which determines the location of the coupling region at different optical states of the system (Schwope et al. 2001). A similar conclusion was reached by G{\"a}nsicke et al. (2006) in a multi-epoch study of the low-state FUV flux of AM~Her, where they calculate a migration of the colatitude WD spot on AM Her by 15$\arcdeg$ in the low state (with respect to the high state). Therefore, it is possible that the HeI stationary components originate from the new threading region perhaps as gas is heated when it impacts the magnetic field lines of the white dwarf, and/or due to interactions between the magnetic field of the red dwarf and the white dwarf. In this case, the gas has a projected velocity around the center of mass of the system thus appearing as a low velocity component in the Doppler map. Similarly, the part of the H$\alpha$ emission that corresponds to the source of the HeI emission is the stationary component (around the center of mass of the system).

It is interesting to point out that stationary components in the optical have been reported in the outburst spectrum of the dwarf novae SS Cyg and IP Peg (Steeghs et al. 1996). In particular IP Peg is an eclipsing CV, providing the opportunity to study the emission sites of the various spectral lines confining their possible origin. In their analysis Steeghs et al. (1996) argue that the stationary component is due to a compact source near the system center of mass; in-eclipse residual flux (from the stationary component) indicate a significant vertical extend of the emitting source. The authors attribute this component to material moving along slingshot prominences (also see Collier Cameron $\&$ Woods 1992) which are kept in place in stable magnetic loops between the L1 point and the center of mass of the system and is pulled towards the center of mass by the gravitational field of the white dwarf itself. In magnetic CVs interactions between the magnetic fields of the two stars are a complicated problem especially in the region around the L1 point where the magnetic fields are of approximate equal strengths ($\sim$kG) and a magnetic ``soup'' is dynamically easily formed and maintained. Prominence-like magnetic structures originating on the donor star can be held in place in such configurations aided by the magnetic field of the WD. Such ``super prominences'' can have a significant vertical velocity components (outside the orbital plane) thus contributing to the low-velocity budget of the emission line. Therefore, in AM Her the low velocity component can originate from activity on the donor star. The observed velocity corresponds to material moving along closed prominence-like loops emerging from the donor star kept in place by fast rotation and magnetic field interactions between the two stars.
 A similar interpretation was proposed for the high-state narrow NV emission component of the AM Her HST/GHRS spectra, which is located between L1 and the center of mass of the system and was found to corrotate with the binary (G{\"a}nsicke et al. 1998). Finally, the H$\alpha$ doppler map of figure~\ref{dopmapHa} also indicates that ``central'' emission line has contributions from the inner hemisphere of the donor star at L1 and extends to the white dwarf, with a component coinciding with the main accretion pole on the white dwarf. The central and right panels of figure~\ref{dopmapHa} clearly demonstrate that the right side of the secondary's inner hemisphere is more strongly irradiated than the left side, consistent with the asymmetric pattern which results from shielding of part of the surface of the donor star by the accretion column.

\subsubsection{H$\alpha$ Satellites}

The two satellite line components map into two lobe-like structures located left and right of L1, with velocities reaching 500km/s. The location of these structures in the Doppler map is completely different from that expected for an accretion stream (either leaving the L1 point and/or following the magnetic field lines). Complete lack of contribution on the lower-left quadrant of the doppler map (no fan-like emission nor ``diffuse'' emission) argues against the presence of a ``classic'' magnetic accretion column toward one or two poles. Also, the observed patterns do not resemble the ones presented in the synthetic AM Her tomograms of Staude et al. (2004) which take into account a z-component of the velocity field reconstructing the high-state HeII and H$\beta$ tomograms. 

If the coupling region has moved towards the donor star (closer to L1) in the low state, it would be energetically favorable for material to flow toward two poles of the system. Within its rich observational history, AM Her has manifested two-pole accretion modes during its high state, mostly detectable in x-rays (eg. Crosa et al. 1981). Therefore it is logical to examine if the two observed lobes in the H$\alpha$ doppler map correspond to accretion onto two magnetic poles. We located in the literature doppler maps of two-pole accreting magnetic CVs as textbook examples of different accretion modes for AM Her. Perhaps one of the better examples is VV Pup for which Diaz \& Steiner (1994) discuss the source of the different emission regions on its high-state doppler map. Except for the accretion stream, the magnetosphere and the irradiated inner hemisphere of the donor star (which are all placed on the left side of the map and along its y-axis), they argue that the H$\alpha$ line has an additional low-velocity component originating on a ballistic stream closer to the L1 point than the accretion column and/or the magnetically-controlled gas. A possible interpretation for this low-velocity component is that it corresponds to gas flowing along ``closed loops'' at the coupling radius, but the authors suggest caution since no secure conclusion could be reached with their data in hand. Our H$\alpha$ doppler map of AM~Her bears no resemblance to the one in VV Pup nor any other accreting magnetic CV. In the low-state tomogram of figure~\ref{dopmapHa} there is no indication for the presence of a ``normal'' accretion stream resulting from Roche-lobe overflow through L1. Gaensicke et al. (2006) find that the hotspot on the WD surface is present at all low states, albeit with lower temperature, indicating that material should be reaching the white dwarf through wind accretion.  However there is no indication for the presence of a second hot pole on the surface of the white dwarf in the low states of AM Her. If we compare the findings of the two FUV studies of Gaensicke et al. (1998; 2006) for the high and low states of the system, the differences between the high and low state of AM Her lie in the spot size, its colatitude on the white dwarf surface and the temperature of the white dwarf between times of ``normal'' accretion and low-state accretion. In the near-IR (J,H and K) cyclotron emission suggests the presence of low-level accretion onto the primary 13.5 MG pole of the system (Campbell, Harrison $\&$ Kafka 2008) but no secure conclusion can be reached for the presence of a second accreting pole. 

\begin{figure}
\epsscale{1.0}
\plotone{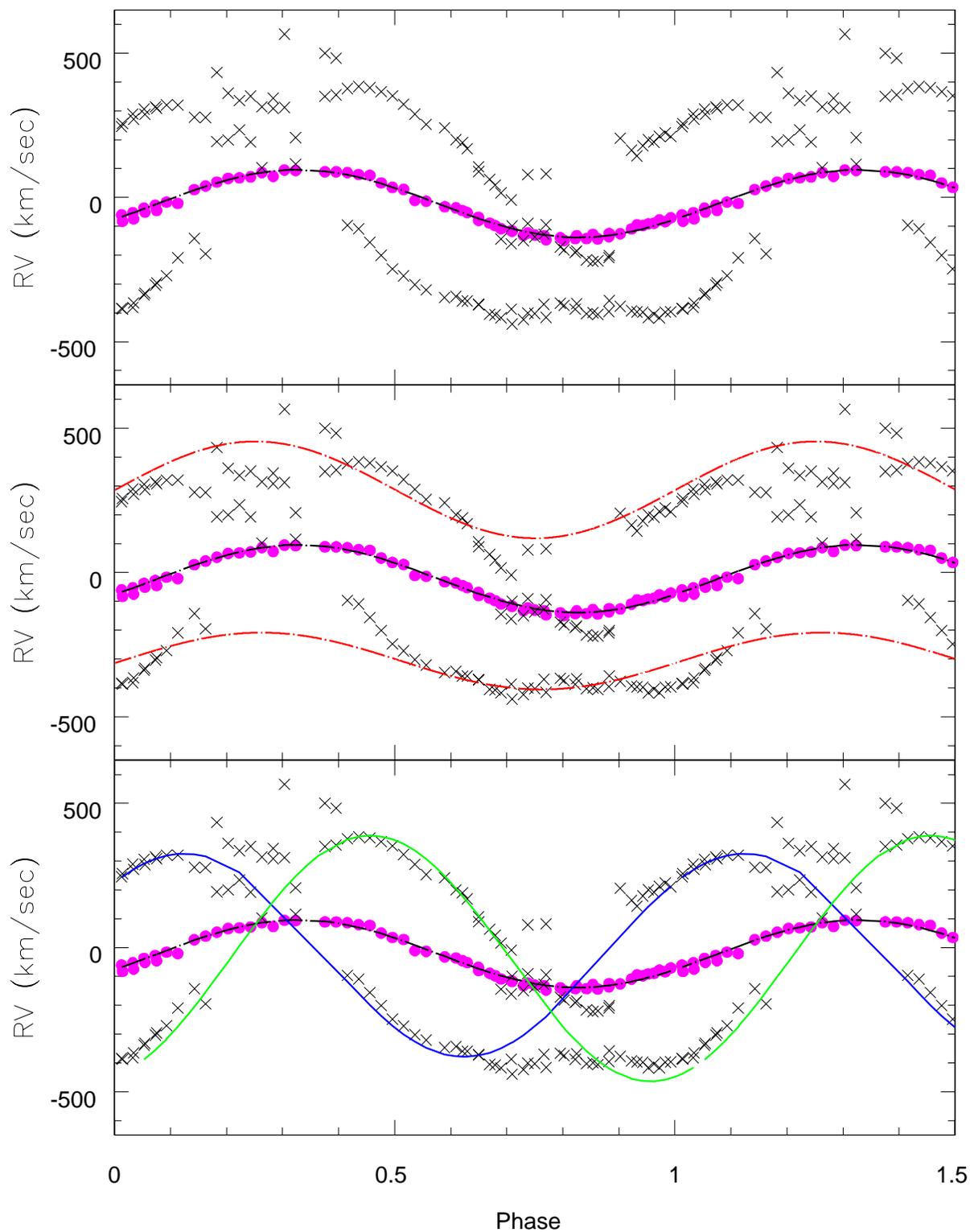}
\caption{Fits of the different line components (central peak and two mail satellites) for the H$\alpha$ line. The derived $\gamma$, K and $\phi$ values are included in table~3. See text for details.\label{rvfits}}
\end{figure}

\begin{figure}
\epsscale{1.0}
\plotone{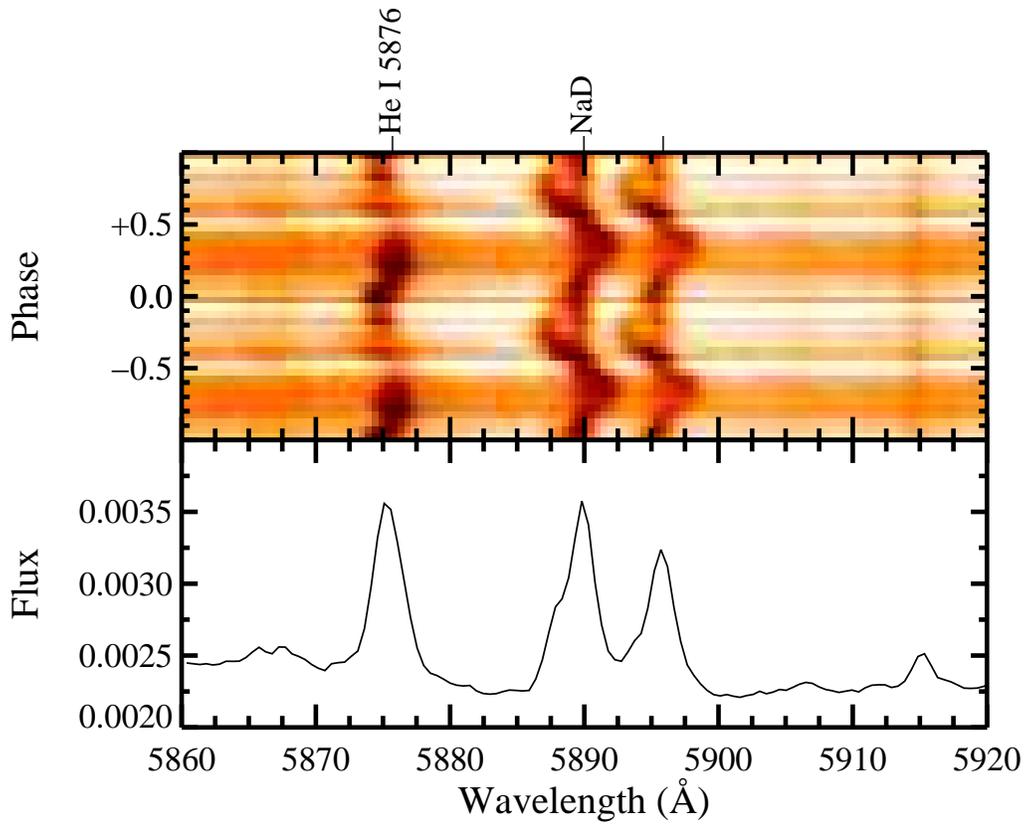}
\caption{Trailed spectra of HeI 5876 $\AA$ and NaD. The stationary component of the NaD line is from the Tucson city lights. The HeI line also appears to have a low velocity, almost stationary component.\label{trailedHeNaDHa}}
\end{figure}

\begin{figure}
\epsscale{1.2}
\includegraphics[angle=-90,scale=.75]{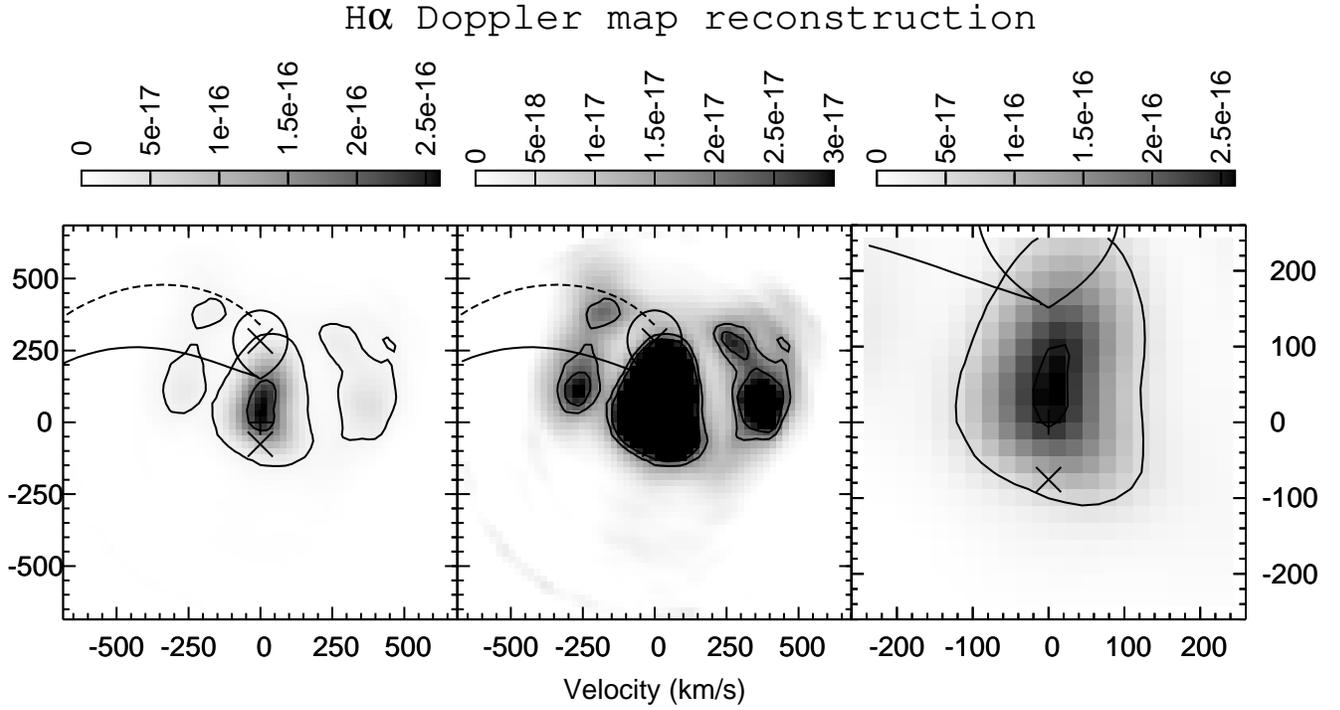}
\caption{H$\alpha$ doppler map. The left and central panels are representations of the system in two intensity scales; the right panel is a zoomed-in version of the region around the center of mass of the system. The intensity scale is presented at the top of each plot.\label{dopmapHa}}
\end{figure}

\begin{figure}
\epsscale{1.0}
\includegraphics[angle=-90,scale=.75]{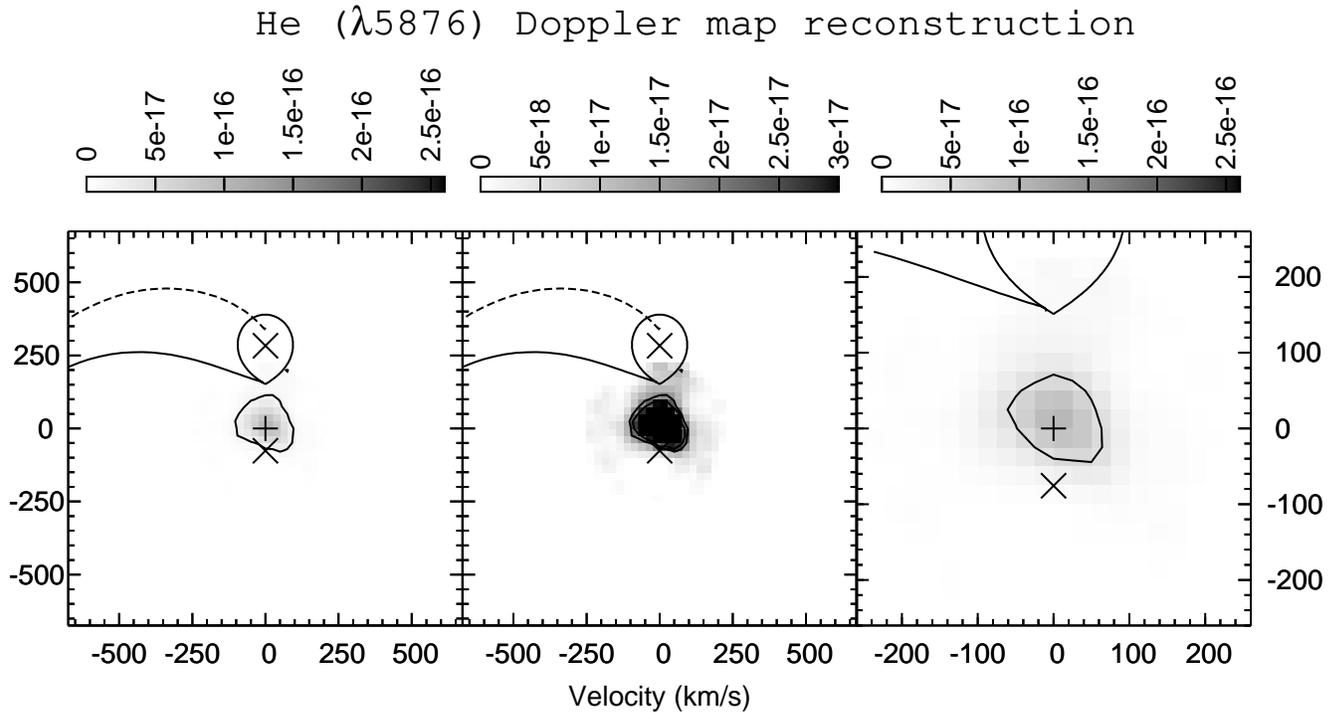}
\caption{Same as figure~\ref{dopmapHa}, for the HeI 5876$\AA$ line \label{dopmapHeI5876}}
\end{figure}

\begin{figure}
\epsscale{0.9}
\includegraphics[angle=-90,scale=.75]{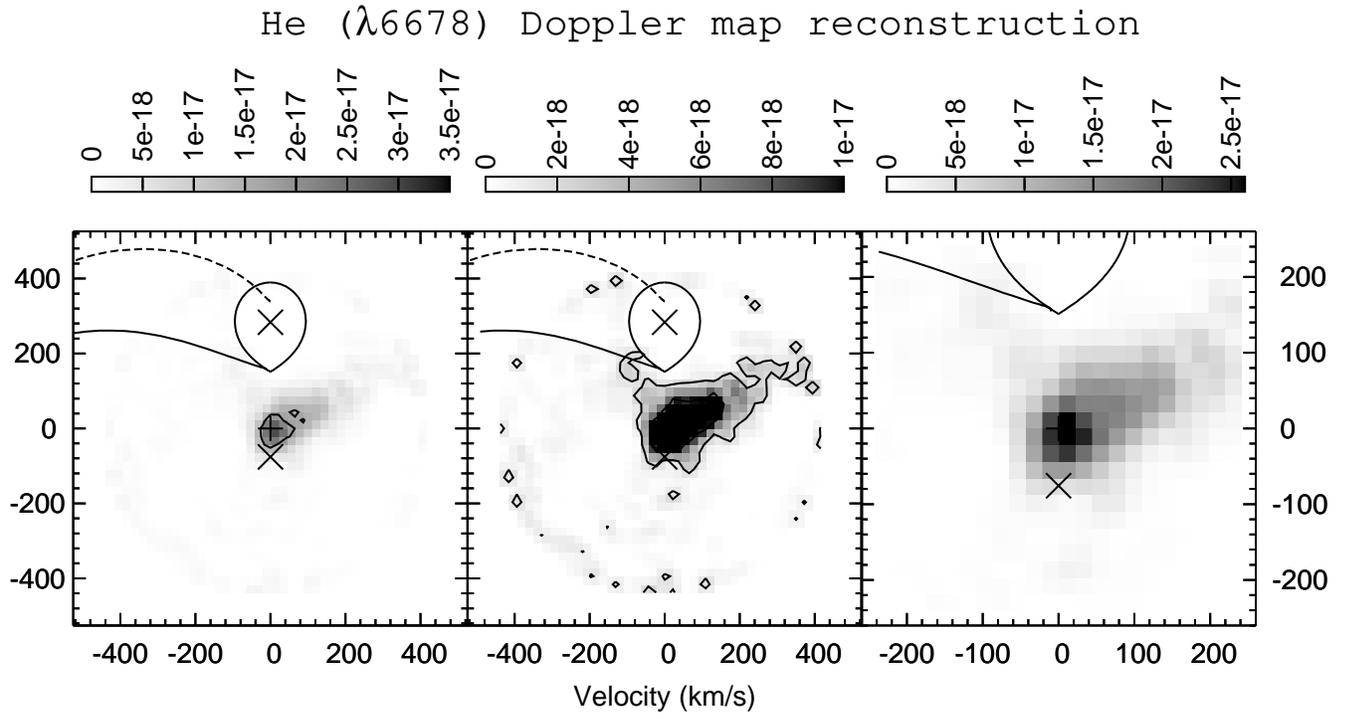}
\caption{Same as figure~\ref{dopmapHa}, for the HeI 6678$\AA$ line\label{dopmapHeI6678}}
\end{figure}

In our ephemeris, the self-eclipse of the accreting pole by the WD happens between orbital phases 0.3 and 0.7 (Campbell et al. 2008) during which both H$\alpha$ satellites are present (figures~\ref{trailedHa} and \ref{rvfits}). This argues against the satellites being due to accretion onto a second pole or even due to accretion onto the main pole of the AM Her white dwarf since none of the satellites is even partially eclipsed between phases 0.3-0.7. 
There is always a chance that accretion takes place in a mode that is unexplored. A possible scenario could favor material escaping the surface of the donor star (anywhere outside the L1 point) following magnetic field lines leading onto the white dwarf itself. If true, we can not identify its footprint on the doppler map of the H$\alpha$ line since the doppler map does not provide clues that material is being directed onto the magnetic pole(s) of the white dwarf.

\begin{figure}
\epsscale{0.9}
\includegraphics[angle=-90,scale=.80]{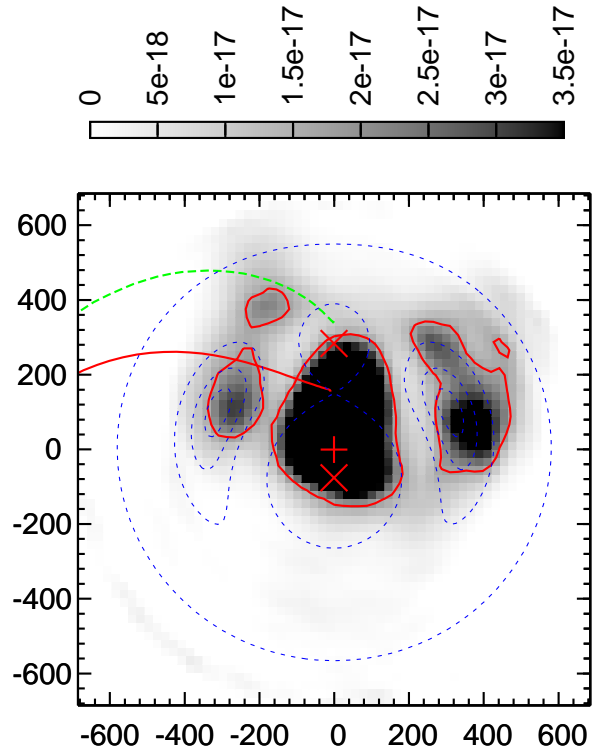}
\caption{Reproduction of the doppler map of figure~\ref{dopmapHa}, with the equipotential surfaces of AM Her overplotted (dashed lines). The plot demonstrates that, kinematically, the two bright satellites coincide with the positions of the L4 and L5 points of the binary.\label{potential}}
\end{figure}

It seems that the H$\alpha$ satellites are a different and persistent feature of the low-state spectra of AM Her.
In papers I and II, we argued that the two satellites represent magnetically confined gas motions in large, long-lived loop prominences on the secondary star. The new data in hand are consistent with this interpretation. These satellites have been present at nearly the same amplitude and phasing (i.e., at the same location on the Doppler map) for the multiple epochs of low-state spectroscopy of AM Her - in Paper I, Paper II, and this paper. As seen in Figure~\ref{rvfits} the RV curves resemble partial sinusoids which are difficult to characterize. 
 The two intersecting sinusoids shown in the bottom panel of Fig~\ref{rvfits} are mapped as the two side lobes in the Doppler tomogram of figure~\ref{dopmapHa}. These two lobes are not well concentrated because the two intersecting sinusoids are incomplete in the sense that some portions of the sinusoids are missing, and some of the data is non-conforming. Understanding how the representation of two non-intersection sinusoids (middle panel of Figure~~\ref{rvfits}) would map into the velocity map is more complicated because sinusoidal RVs with large $\gamma$ velocities are not at all concentrated to a specific location in the Doppler map (see Kaitchuck et al. 1994 and Schwope et al. 1997 for discussions of this behavior). The simplest explanation for the satellites would require two main loops which are spatially isolated and kinematically independent. Their appearance in the doppler map (figure~\ref{dopmapHa}) resemble an incomplete ring around the donor star; this can be produced by material following closed magnetic paths emerging from the donor star. The location of the two loops in the Doppler map suggests that material emerging from the donor star could be trapped in the L4 and L5 Lagrangian points of the system as demonstrated in figure~\ref{potential}; even their elongation in the Doppler maps are in the
   same direction as the equipotential contours. However, L4 and L5
   are stable equilibrium locations only for q$\lesssim$0.04, requiring
   a brown dwarf for the secondary star (M$_{2}$$\lesssim$0.08M$_{\sun}$). We support instead that streaming (prominence-like) motions from the secondary star dominate the location of the satellites in the Doppler map.

\subsubsection{System Parameters}

\begin{figure}
\epsscale{0.9}
\includegraphics[angle=-90,scale=.75]{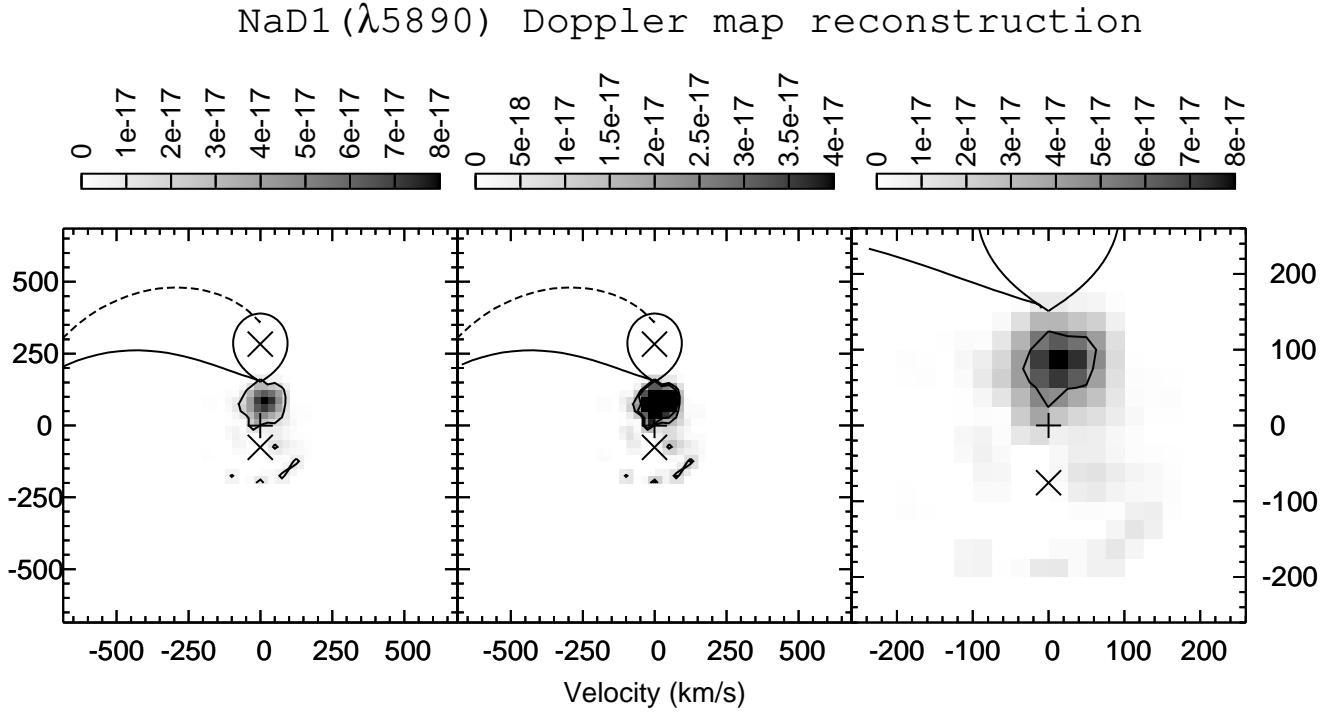}
\caption{Doppler map of the NaD 5890$\AA$ line, for i=60$\degr$ and q=0.267. The majority of the emission seems to originate from a point between L1 and the center of mass of the system. The stationary component due to Tucson city lights has been removed.\label{NaD1a}}
\end{figure}

The above discussion depends heavily on the relative location of L1 and of the center of mass of the system with respect to the observed features on the velocity map.  The adopted inclination {\it i} and mass ratio (q=M$_{2}$/M$_{WD}$) are listed in table~2. According to those values, the majority of the low velocity emission in H$\alpha$ and in the HeI lines originates from a source between L1 and the center of mass of the system, allowing for little contribution from the irradiated inner hemisphere of the donor star. Perhaps this should be expected in the low state of AM Her due to reduced heating from the accreting pole. In figure~\ref{NaD1a} we present the velocity map of the NaD~5890$\AA$ line for the chosen values of table~2 (a similar map results from the NaD2 line, which is not shown here). The figure demonstrates that the source of the line is located between L1 and the center of mass of the system. We see no evidence that this feature is due to irradiation. 

\begin{figure}
\epsscale{0.7}
\includegraphics[angle=-90,scale=.75]{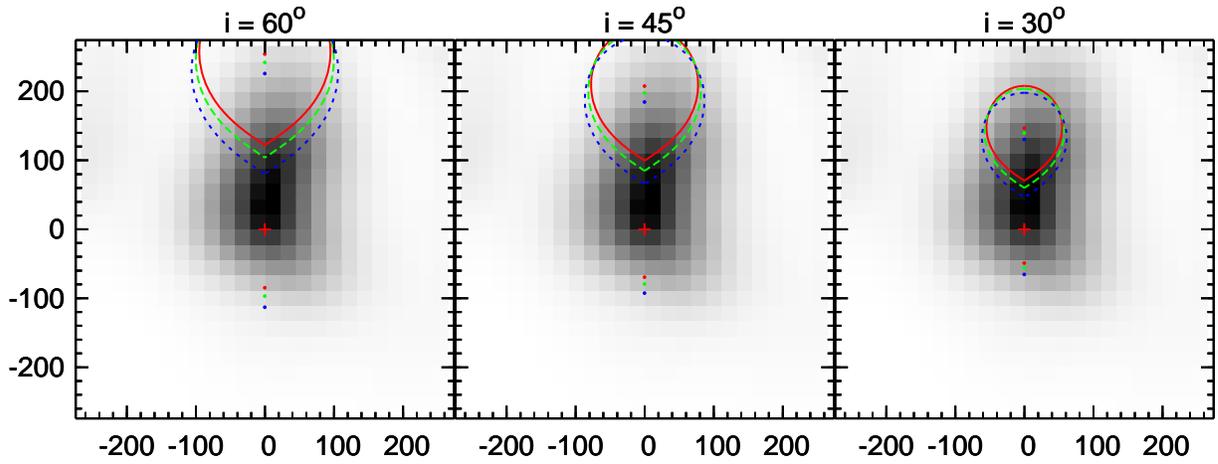}
\caption{Doppler maps of the H$\alpha$ emission line with different values of system inclination {\it i} and mass ratio q (q=M$_{2}$/M$_{WD}$). From left to right: {\it i}=60$\degr$, 45$\degr$, 30$\degr$. For each panel, q=0.33 (red), 0.45 (green) and 0.5 (blue). See text for discussion.\label{testHa1}}
\end{figure}

\begin{figure}
\includegraphics[angle=-90,scale=.75]{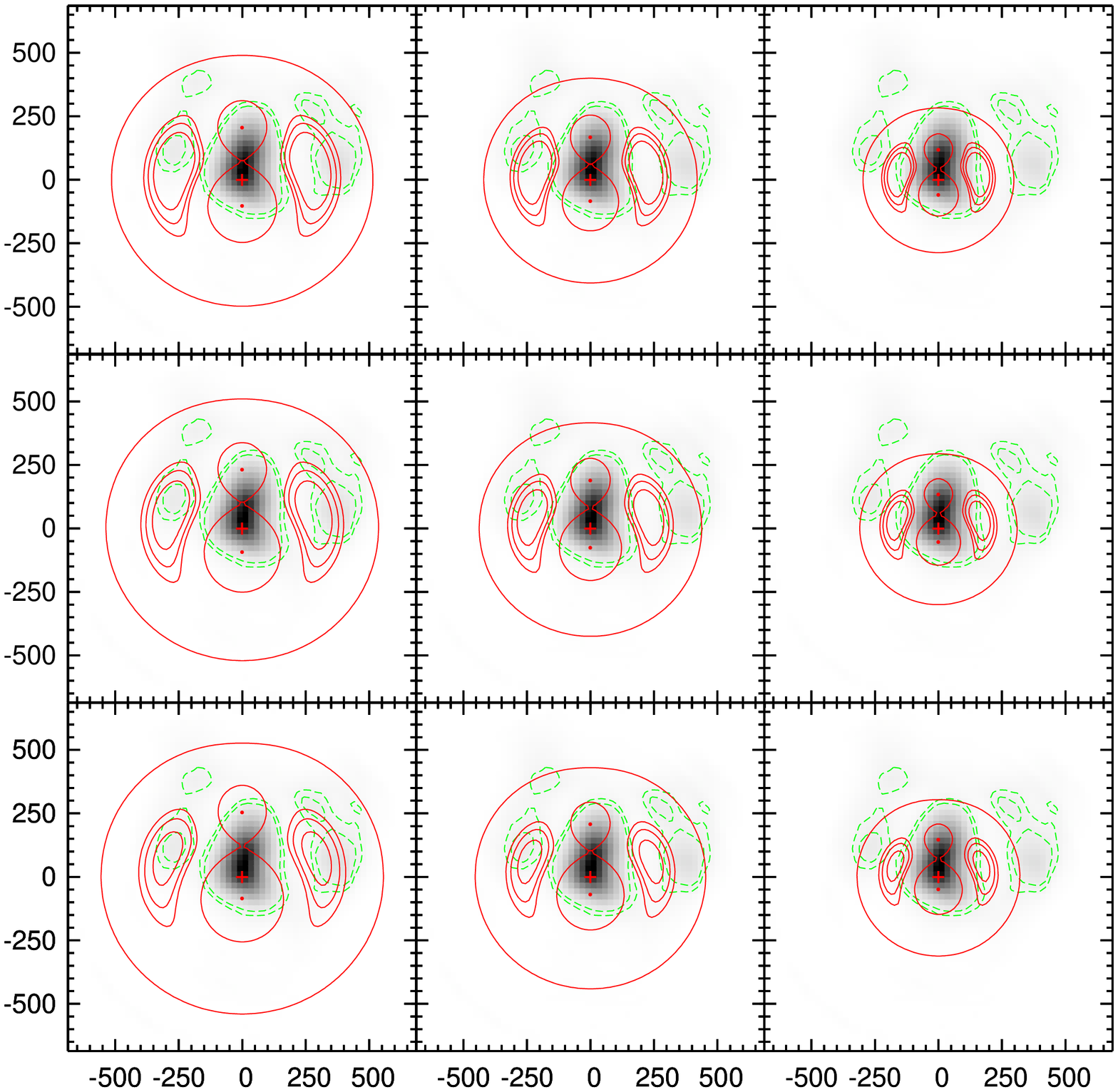}
\caption{Same as figure~\ref{testHa1}, with the the equipotential surfaces of AM Her overplotted (as in figure~\ref{potential}). From left to right, {\it i}=60$\degr$, 45$\degr$, 30$\degr$ and from bottom to top q=0.33, 0.45 and 0.5. The green-dashed line corresponds to the equal intensity contours of the middle panel of f~\ref{dopmapHa}. (See text for discussion.)\label{testHa2}}
\end{figure}

\begin{figure}
\includegraphics[angle=-90,scale=.60]{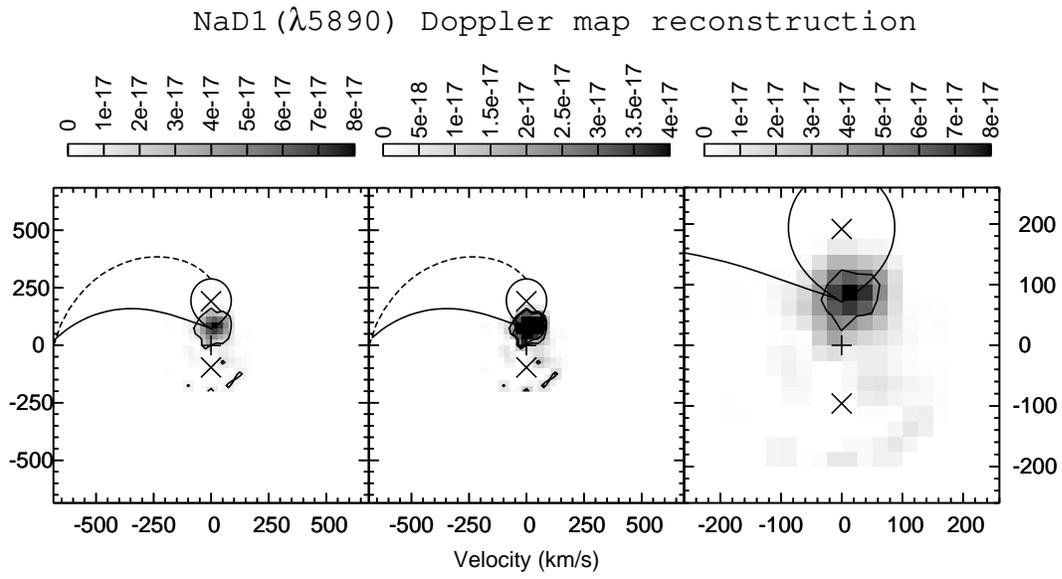}
\caption{Doppler map of the NaD 5890$\AA$ line, for i=45$\degr$ and q=0.45. (See discussion in text).\label{NaD1b}}
\end{figure}

We explored the ({\it i},q) parameter space, in order to see if it is possible to find a geometry in which the low-velocity component could naturally be explained in terms of irradiation of the mass-donor star (i.e., emission from near the L1 point)  as required for an irradiation mechanism. In figure~\ref{testHa1} we present the H$\alpha$ doppler map (of figure~\ref{dopmapHa}, left) with different values of system inclination {\it i} and mass ratios q superimposed. From left to right, we use {\it i}=60$\degr$, 45$\degr$ and 30$\degr$ for mass ratios of 0.33, 0.4 and 0.5 (top to bottom). This figure indicates that there are pairs of (i,q) that allow for {\it part} of the low velocity emission component to originate close to L1. A different representation is presented in figure~\ref{testHa2} which overplots the system's equipotential surfaces (including L4 and L5) for different values of {\it i} and q. In the plot we include contours outlining the emission components in the intensity scale of the middle panel of figure~\ref{dopmapHa}.  From this second plot - if we take into account out irradiation scenario, solutions with {\it i}$<$45$\degr$ and q$>$0.5 are unlikely because they would imply a roughly uniform emission from the surface of the secondary star. 
Nevertheless, these solutions could not be discarded, as many polars 
 do show H$\alpha$ emission while in the low state from the active chromosphere of the secondary star at all phases.
 Interestingly if we require for the NaD1 emission to be placed on the irradiated inner hemisphere of the donor star, we have the map of figure~\ref{NaD1b} for i=45$\degr$ and q=0.45, however this solution is not unique either.  Nevertheless, our interpretation of the satellites and low-velocity H$\alpha$ emission components does not depend on the adopted (and rather uncertain) binary geometry.

\section{Final Remarks}

In this paper we present new high time resolution optical spectra of the magnetic CV prototype AM Herculis during its 2006 low state. We compute Doppler maps to examine the origin of various emission lines and we identify three main emitting sources in the system. In particular, the stationary components of the HeI and H$\alpha$ likely arises from gas coupling onto the magnetosphere of the donor star; this seems to have migrated closer to the L1 point likely due to the significantly reduced mass accretion rate in the low states. The two high velocity H$\alpha$ components likely represent gas streaming along
 prominence-like structures kept in place by interactions of the magnetic dipoles of the two stars. 

It is not clear from our data whether the RV curves of the H-alpha
  satellite components form two crossing sine curves, or two sine-like
     curves offset from one another. Under the latter characterization
     we argued in Papers I and II that the satellites originate from streaming
     motions in large loop prominences tied to the secondary star.  If the
     two RV curves cross, then it seems that other possibilities are
 allowed,
     but that loop prominences associated with the secondary star are not
     ruled out.  It seems likely that the motion of gas between the two
     stars during the low state is along magnetic field lines associated
     with {\it both} stars, in a complex pattern giving rise to several
     partial sinusoids in the trailed spectra.  Detailed modeling will be
 needed to unravel this behavior.  The fact that the features are stable and
     repeatable will make this future effort worthwhile.

We are making our first steps in understanding the complex nature of magnetic fields in two corotating magnetically interacting stars, but with AM Her we have touched the tip of the iceberg. Four magnetic CVs\footnote{In addition to AM Her the magnetic CVs of the group are ST LMi (Kafka et al. 2007), VV Pup (Mason E. et al. 2008) and EF Eri (Kafka et al. 2008).} have been found so far to exhibit additional low-state H$\alpha$ components (including AM Her). It is likely that such structures are present in all magnetic CVs but are masked by the regular accretion stream in the high state. Perhaps through the H$\alpha$ satellites we witness a new mode of magnetic accretion toward the white dwarf through magnetic field interactions of the two stars (as opposed to Roche-lobe overflow). It has been already suggested that magnetic braking may not be important in magnetic CVs of all orbital periods (see for example Li et al. 1994). It is very likely that the observed structures represent an alternative angular momentum loss mechanism for magnetic CVs (e.g. Norton et al. 2008 and references therein). For magnetic braking to be efficient in single stars, magnetic field strengths of the order of 1000G suffice. To date, AM Her (and the present study) is the best example we have of those structures, therefore can provide the testbed for relevant models and a lower limit of the effect of such magnetic structures on CV evolution. It is beyond the scope of this paper to present a model of these structures. Using various techniques (and higher resolution data) we aspire to address (and eventually resolve) the origin of the observed low state structures and their consequence to the evolution of such systems.

\acknowledgments
We would like to thank our anonymous referee for a careful review of the manuscript. We acknowledge with thanks the variable star observations from the AAVSO International Database contributed by observers worldwide and used in this research.

Facilities: \facility{MMT}

\clearpage

\clearpage

\begin{deluxetable}{cccr}
\tablecaption{Parameters of Radial Velocity Curve Fits of the Emission lines}
\tablewidth{0pt}
\tablehead{
\colhead{line} & \colhead{$\gamma$ (km/sec)} & \colhead{K (km/sec)} & \colhead{$\phi_{0}$}}
\startdata
 H$\alpha$ center      &  -23 $\pm$ 1 &  117 $\pm$ 2 & 0.077 $\pm$ 0.002 \\
H$\alpha$ satellite 1 &  -26 $\pm$ 4 &  352 $\pm$12 & 0.874 $\pm$ 0.007 \\
H$\alpha$ satellite 2 &  -38 $\pm$ 3 &  426 $\pm$12 & 0.167 $\pm$ 0.005 \\
HeI 5876$\AA$         &  -26 $\pm$ 1 &   36 $\pm$ 4 & 0.062 $\pm$ 0.017 \\ 
HeI 6678$\AA$         &  -23 $\pm$ 1 &   83 $\pm$ 4 & 0.030 $\pm$ 0.006\\
NaD1 (5890$\AA$)     &  -23 $\pm$ 1 &   91 $\pm$ 3 & 0.032 $\pm$ 0.005 \\
NaD2 (5896$\AA$)     &  -23 $\pm$ 1 &   90 $\pm$ 4 & 0.028 $\pm$ 0.006 \\
\enddata
\end{deluxetable}

\begin{deluxetable}{ccc}
\tablecaption{Model Parameters for the Doppler maps}
\tablewidth{0pt}
\tablehead{\colhead{parameter} & \colhead{value} & \colhead{reference}}
\startdata
i & 60\degr & Watson et al. (2003) \\
M$_{WD}$ & 0.75 M$_{sun}$   & G{\"a}nsicke et al. (2006) \\
M$_{2}$ &  0.20 M$_{sun}$   & Allen (2000)  \\
P$_{orb}$ & 0.1289d & Kafka et al. (2005a)\\
\enddata

\end{deluxetable}


\end{document}